\documentclass[preprint,preprintnumbers,amsmath,amssymb]{revtex4}

\usepackage{graphicx}% Include figure files
\usepackage{bm}% bold math

\begin{document}

%\preprint{APS/123-QED}

\title{Macroscopic Invisibility Cloak for Visible Light}

\author{Baile Zhang$^{1,2}$}
\author{Yuan Luo$^{1,2}$}
\author{Xiaogang Liu$^1$}
\author{George Barbastathis$^{1,2}$}
\email{gbarb@mit.edu} \affiliation{$^1$Singapore-MIT Alliance for
Research and Technology (SMART) Centre, Singapore 117543, Singapore.\\
$^2$Department of Mechanical Engineering, Massachusetts Institute of
Technology, Cambridge, Massachusetts 02139, USA. }


\begin{abstract}
Invisibility cloaks, a subject that usually occurs in science
fiction and myths, have attracted wide interest recently because of
their possible realization. The biggest challenge to true
invisibility is known to be the cloaking of a macroscopic object in
the broad range of wavelengths visible to the human eye. Here we
experimentally solve this problem by incorporating the principle of
transformation optics into a conventional optical lens fabrication
with low-cost materials and simple manufacturing techniques. A
transparent cloak made of two pieces of calcite is created. This
cloak is able to conceal a macroscopic object with a maximum height
of 2 mm, larger than 3500 free-space-wavelength, inside a
transparent liquid environment. Its working bandwidth encompassing
red, green and blue light is also demonstrated.
\end{abstract}

\maketitle

Among various devices conceptualized in the emerging field of
transformation optics~\cite{huanyang_review}, the most attractive
one  might be the cloak of invisibility, which can render an object
invisible by precisely guiding the flow of light around the object,
as if the object does not
exist~\cite{leonhardt,pendry,leonhardt_broad,li_carpet}. The
underlying mechanism stems from the formal invariance of Maxwell's
equations: a coordinate transformation does not change the form of
Maxwell's equations, but only changes the constitutive parameters
and field values. The hidden object is rendered invisible simply
because it is out of the transformed electromagnetic space. Great
interest in realizing invisibility in practice has fueled massive
research efforts from microwave to optical
frequencies~\cite{schurig,smolyaninov_SPP,ruopeng,valentine,gabrielli,park,smolyaninov_waveguide,huifeng,ergin}.

Significant progress has been made during the exploration of
invisibility cloak. The first theoretical model of a
transformation-based cloak~\cite{pendry} required extreme values of
the constitutive parameters of materials used and can only work
within a very narrow frequency band~\cite{pendry,baile_rainbow}.
Schurig \textit{et al.} overcame the  first flaw by using simplified
constitutive parameters at microwaves based on metamaterial
technologies~\cite{schurig}. To bypass the bandwidth limitation and
push the working frequencies into the optical spectrum, it has been
proposed that an object sitting on a flat ground plane can be made
invisible under a fully dielectric gradient-refractive-index
``carpet'' cloak generated by quasiconformal
mapping~\cite{li_carpet}. This carpet cloak model has led to a lot
of subsequent experiments in both microwave~\cite{ruopeng,huifeng}
and infrared frequencies~\cite{valentine,gabrielli,park,ergin}.
However, a serious limitation of carpet cloaks was recently pointed
out: the quasiconformal mapping strategy will generally lead to a
lateral shift of the scattered wave, whose value is comparable to
the height of the hidden object, making the object
detectable~\cite{baile_lateral_shift}. Furthermore, all previous
experiments of invisibility in the optical spectrum, from
infrared~\cite{valentine,gabrielli,park,ergin} to
visible~\cite{smolyaninov_SPP,smolyaninov_waveguide} frequencies,
were demonstrated under a microscope, hiding objects with sizes
ranging from the order of 1
wavelength~\cite{smolyaninov_SPP,valentine,gabrielli,park,ergin} to
approximately 100 wavelengths~\cite{smolyaninov_waveguide}. How to
``see'' the invisibility effect with the naked eye,  i.e. to cloak a
macroscopic object in visible light, is still a crucial challenge.
In addition, most previous optical
cloaks~\cite{smolyaninov_SPP,valentine,gabrielli,park,ergin}
required complicated nano- or microfabrication, where the cloak, the
object to be hidden, and the surrounding medium serving as the
background were all fabricated in one structure. Thus they could not
be easily separated from their embedded structures and transferred
elsewhere to cloak other objects. These limitations, such as
detectability, inadequate capacity to hide a large object, and
nonportability, must be addressed before an optical cloak becomes
practical.

The above limitations boil down to two difficulties in the
fabrication of cloak materials---anisotropy and inhomogeneity. The
previously proposed quasiconformal mapping strategy attempted to
solve anisotropy in order to facilitate metamaterial
implementation~\cite{li_carpet}. However, in conventional optical
lens fabrication, the inhomogeneity is more difficult to implement
than anisotropy. While there is still a long way to go before
achieving practical applications of nanobuilt metamaterials, the
transformation principle can be directly incorporated into
conventional optical lens fabrication, if the difficulty of
inhomogeneity can be lifted. Here, we report the first realization
of macroscopic invisibility cloaking at broadband visible
wavelengths by applying transformation optics design into
conventional optical lens fabrication. The cloak is implemented with
calcite, a common anisotropic optical material. Since calcite itself
is transparent in visible light, the concern of energy loss for
metamaterials does not exist in this case, nor the bandwidth
limitation associated with metal ingredients~\cite{huanyang_review}.
To our knowledge, in our experiment invisibility was demonstrated
for the first time by ``seeing'' through the cloak directly, i.e.,
by placing a target object behind the cloak, illuminating the system
with visible light, and showing that there is no evidence of the
cloaked object in the image of the target object. It is, therefore,
closest to the idealized concept of a cloak---being invisible to the
eye.

\begin{figure}
\begin{centering}
\includegraphics[width=1\columnwidth,draft=false]{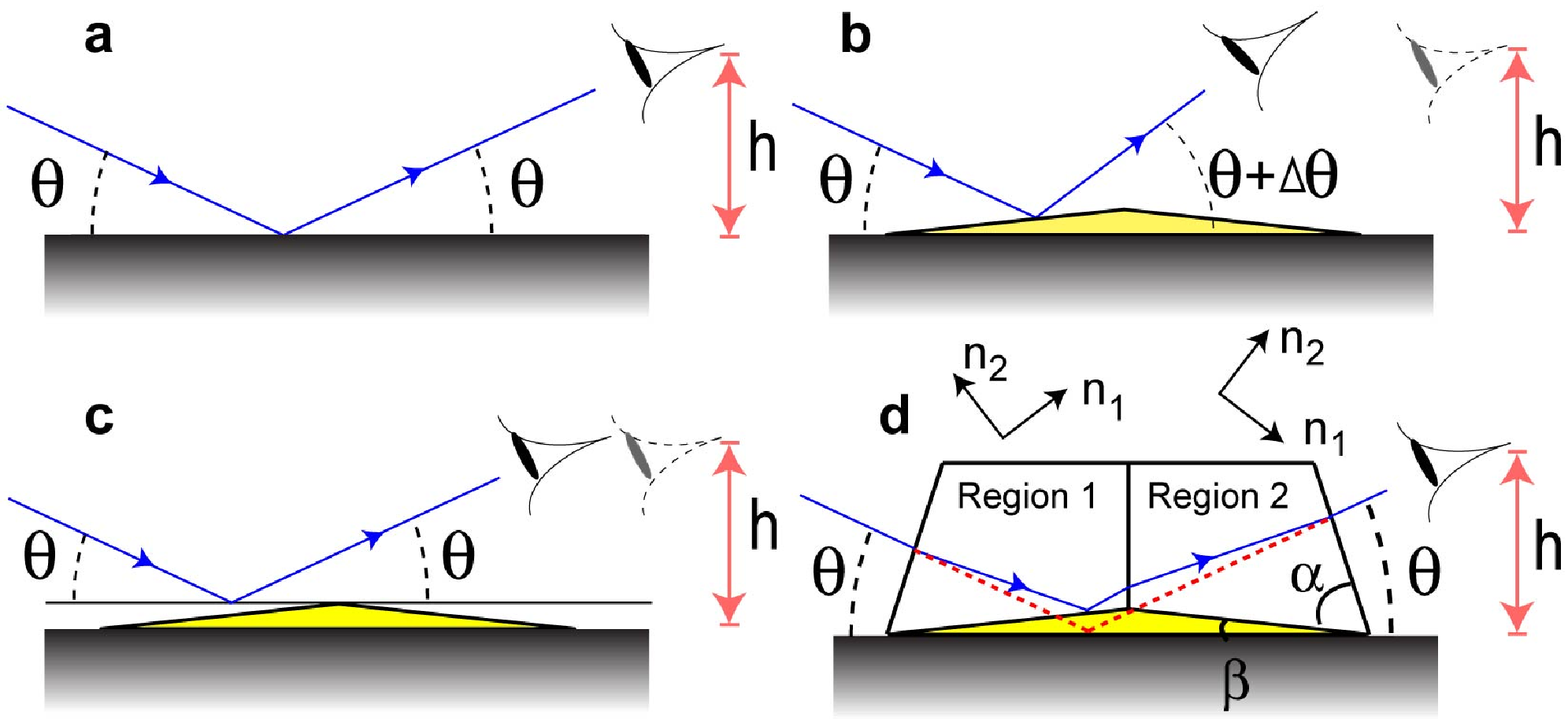}% Here is how to import EPS art
\caption{\label{fig:rays2}  (a) A light ray is incident on a
flat ground plane and reflected back with the same angle.
(b) When an object is sitting on the ground plane, the
reflected ray changes its angle. (c) When another flat
ground plane is put on top of the object, the reflected ray restores
its angle but suffers a lateral shift. (d) When a
transformation-based anisotropic cloak is covering the object, the
reflected ray restores both its angle and position. The anisotropic
medium has two principal refractive indexes $n_1$ and $n_2$ along
two orthogonal directions. The observer in all cases is assumed to
have a fixed height of $h$. In (b) and (c), the
original position of the observer is indicated with a dotted eye.}
\end{centering}
\end{figure}

To explain our cloak design, let us first consider why an object is
visible. Figure~\ref{fig:rays2}(a) shows a ray of light incident on
the ground plane with an angle $\theta$. As the observer can see,
when there is nothing on top of the ground plane, the light will
just be mirror reflected with the same angle $\theta$ from the
ground plane. When an object is placed on top of the ground plane,
like the triangle in Fig.~\ref{fig:rays2}(b), the reflected light
will change its reflection angle. This change can be easily noticed
by the observer. The most convenient way to restore the angle of the
reflected light is to put another flat ground plane on top of the
object directly, as shown in Fig.~\ref{fig:rays2}(c). However,
although the angle has been restored, the reflected light
experiences a noticeable lateral shift, which unveils the existence
of the object. It is, therefore, necessary to restore both the angle
and the position of the reflected light in order to render the
object invisible under a cloak. A homogeneous cloak of a triangular
shape made of uniaxial medium is possible if the electromagnetic
space above the triangular object is squeezed upwards uniformly
inside another larger triangle~\cite{xisheng}, since the optical
path is preserved in this transformation. For material savings
without affecting the essential cloaking function, we can truncate
the cloak to a trapezoid with a small triangle etched away from the
bottom. The final structure is shown in Fig.~\ref{fig:rays2}(d). It
is necessary that the material in the left and right regions should
have two principal refractive indexes, $n_1$ and $n_2$, as indicated
in the figure~\cite{xisheng}. The background should have a
refractive index $n$ matched to a value between $n_1$ and
$n_2$~\cite{zhang10} though using air as background is also
possible~\cite{chen10}. Ray tracing in Fig.~\ref{fig:rays2}(d)
clearly shows that, within the range limited by the truncated size
of the cloak, an arbitrary incident ray can be guided ideally and
emitted with the same angle and from the same position as in
Fig.~\ref{fig:rays2}(a). This is because the anisotropic cloak is
equivalent to a squeezed electromagnetic space, where the light is
``deceived'' as if it were propagating in the empty space of
Fig.~\ref{fig:rays2}(a) with nothing on top of the ground plane.

To implement this anisotropic cloaking principle easily at optical
frequencies, we chose a two--dimensional (2D) geometry with light
polarized such that the magnetic field is parallel to the horizontal
reference plane. Let us first consider the following
transformation~\cite{xisheng} from the original coordinates
($x$,$y$) into a new set of coordinates ($x'$,$y'$)
\begin{eqnarray}
\left\{\begin{array}{c} x'=x;
\\
y' = \kappa y + \tau (a - |x|),
\end{array} \right.
\end{eqnarray}
where $\kappa = (\tan (\alpha + \beta)- \tan \beta) / \tan
(\alpha+\beta)$ and $\tau = \tan \beta$.

Through this transformation, we can obtain the constitutive
parameters for the material in the first quadrant as
\begin{eqnarray}
\overline{\overline{\epsilon}}\:{}' & =  & \frac{\overline{\overline J} \cdot \overline{\overline{J}}^{T}}{|\overline{\overline J}|} = \left( \begin{array}{cc} 1/\kappa & -\tau /\kappa   \\
-\tau /\kappa & \kappa +\tau^2/\kappa
\end{array} \right); \\
\mu' & = & \frac{1}{\kappa},
\end{eqnarray}
where $\overline{\overline J}$ is the Jacobian of the
transformation. The material in the second quadrant has the same
parameters except for a mirror reflection with respect to the vertical plane.

To make this work with a nonmagnetic material, we scale the permittivity tensor such that
\begin{equation}
\overline{\overline{\epsilon}}\:{}'{}_{\text{nm}} = \left( \begin{array}{cc} 1/\kappa^2 & -\tau /\kappa^2\\
-\tau /\kappa^2 & 1 +\tau^2/\kappa^2\\
\end{array} \right).
\end{equation}
This guarantees $\mu'{}_{\text{nm}}=1$. To determine the principal axes of the uniaxial crystal, we diagonalize $\overline{\overline{\epsilon}}\:{}'{}_{\text{nm}}$ and obtain the crystal axis orientation with respect to the vertical plane as
\begin{equation}
\gamma = {1\over 2} \arctan \frac{2 \tau}{\kappa^2 + \tau^2 -1} \label{eq:gamma}.
\end{equation}

To fix the geometry, we let $\alpha = 66^{\circ}$ and $\beta = 6^{\circ}$ and obtain $n_1^2=1.48^2$, $n_2^2=1.66^2$, and $\left|\gamma\right|=37.5^\circ$ for background refractive index equal to 1.57. For green light at wavelength 561nm (where human eyes are most sensitive in common environments~\cite{wyszecki}) and magnetic field oriented parallel to the mirror, this can be realized simply as calcite with its crystal axis oriented according to (\ref{eq:gamma}).

\begin{figure}
\begin{centering}
\includegraphics[width=1\columnwidth,draft=false]{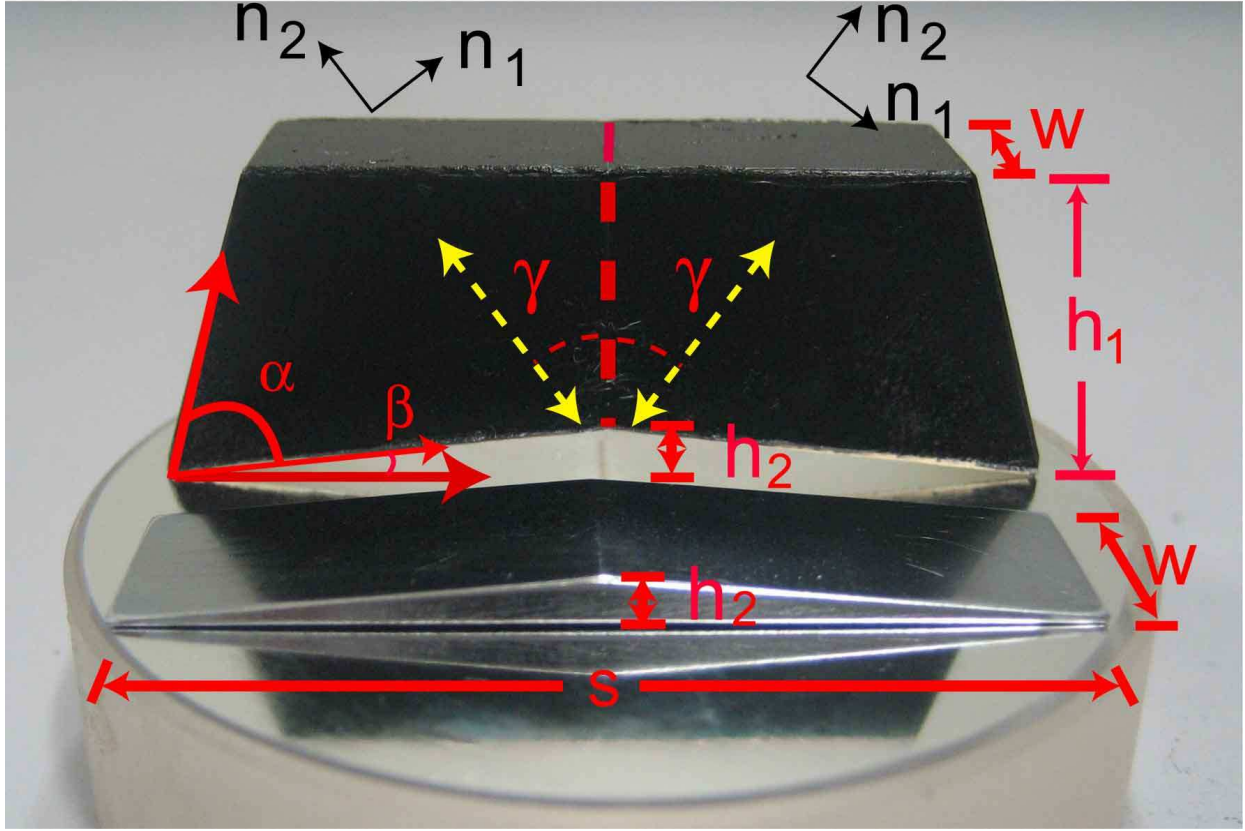}% Here is how to import EPS art
\caption{\label{fig:setup} A transformation-based anisotropic cloak
compared to a steel wedge on top of a mirror. The cloak is made of two
pieces of calcite crystal with specific orientations of the optical
axis indicated by the yellow dotted arrows. For green light, with
its magnetic field perpendicular to the optical axis, $n_1 = 1.48$
perpendicular to the optical axis and $n_2 = 1.66$ along the optical
axis. $\alpha=66^{\circ}$, $\beta = 6^{\circ}$, and $\left|\gamma\right| =
37.5^{\circ}$. $w=10$ mm, $h_1 = 14.5$ mm, $h_2 = 2$ mm, and $s=38$
mm.}
\end{centering}
\end{figure}

We fabricated the experimental cloak as shown in
Fig.~\ref{fig:setup}. Because of fabrication errors in the calcite
crystal, angles were accurate up to $15'$, while the length of each
side may have error up to 1 mm. The two pieces of calcite were
back-to-back cemented at the location of the dotted line indicated
in Fig.~\ref{fig:setup} with Norland Optical Adhesive 61, a clear,
colorless, liquid photopolymer that cures when exposed to
ultraviolet light. The front, back, and top surfaces of the cloak
were painted black to absorb scattered light. The bottom was fully
polished and then coated with silver. The left and right surfaces
were fully polished and served as the input and output for the
light, respectively. The object to be hidden was a steel wedge that
fit into the void under the cloak and is also shown in
Fig.~\ref{fig:setup}. Any object smaller than this steel wedge can
be alternatively adopted. Since this design aims at invisibility in
a transparent medium with refractive index close to $n=1.57$, a
value similar to the background refractive index in
\cite{valentine}, it is not subject to the limitations of
delay-bandwidth and delay-loss for cloaking in air~\cite{hashemi}.

\begin{figure}
\begin{centering}
\includegraphics[width=1\columnwidth,draft=false]{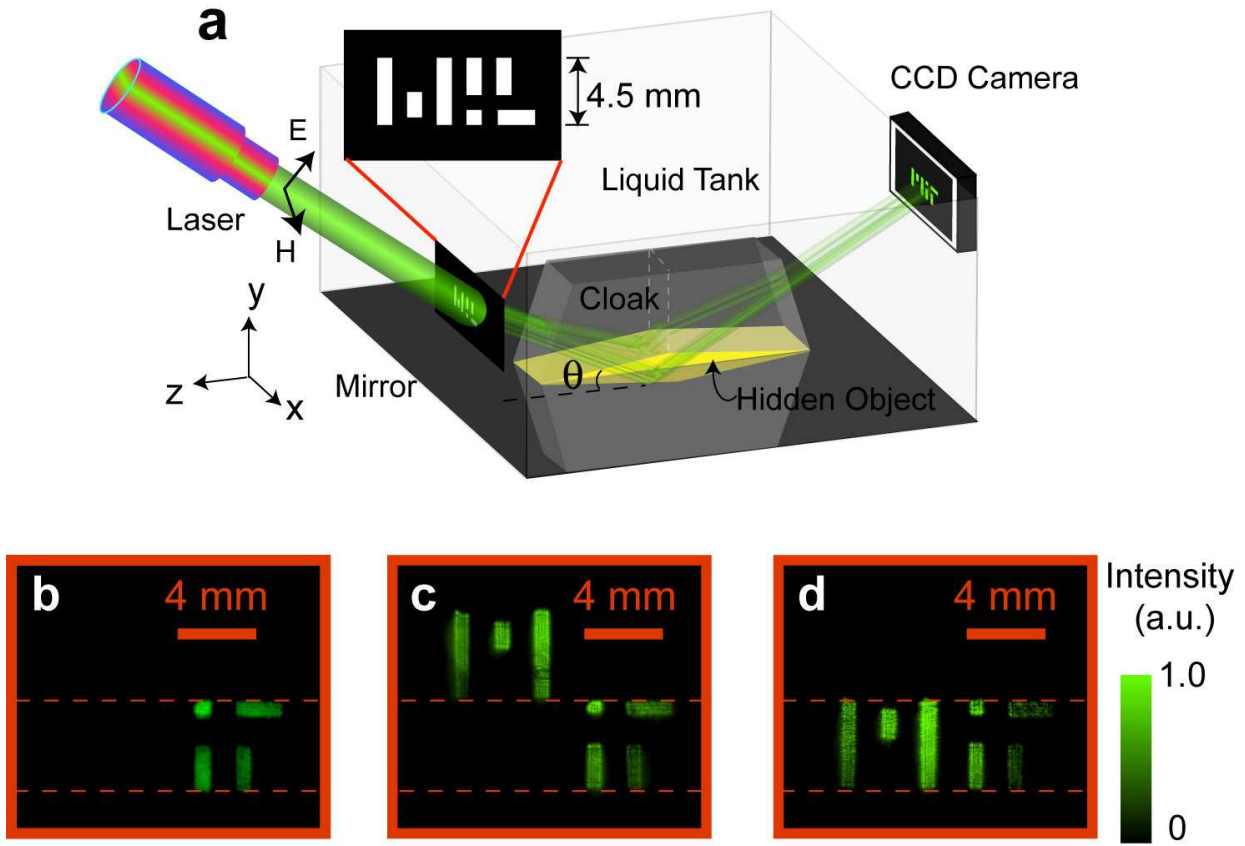}% Here is how to import EPS art
\caption{\label{fig:green} (a) Schematic diagram of the
experimental setup and images captured on the CCD camera
(b) with the wedge only but
without the cloak, (c) with a flat ground plane on top
of the wedge, (d) with the cloak on top of the wedge. }
\end{centering}
\end{figure}

As shown in Fig.~\ref{fig:green}(a), all objects: the cloak, the
wedge to be hidden, and the mirror used as the ground plane, were
immersed in a glass tank filled with a transparent colorless laser
liquid (OZ-Laser IQ, CODE 5610, from Cargille Labs, $n=1.53$
measured at wavelength 598.3 nm). A hollow transmission pattern
reading ``MIT'' was printed on an opaque plastic plate with
thickness of 500 $\mu$m using a stereolithography machine
(Viper$^{\rm TM}$ SLA System). The resin was Accura 60 Plastic. This
plate was inverted and then attached to the left--hand side of the
transparent tank with adhesive tape. Because of the finite thickness
of the plate, the final image of ``MIT'' looked slightly slimmer in
the experiments.

The hollow pattern was then illuminated by a continuous wave (cw)
laser diode at wavelength of 561nm polarized with the magnetic field
parallel to the mirror. The position where the pattern was placed
was carefully chosen such that the light transmitted through the
inverted ``M'' went through the cloak with the hidden wedge
underneath and was reflected at the bottom of the cloak, while the
light through the inverted ``IT'' was reflected on the mirror's
surface directly without touching the cloak and the wedge. A
charge-coupled-device (CCD) camera (Kodak ISS KAI-11002 with pixel
size of 9 $\mu$m) was placed at a distance of approximately 10cm
from the cloak on the right--hand side outside the tank. Since this
homemade cloak had side error up to 1 mm, in the experiment we
adjusted the height of the cloak by inserting 100$\mu$m--thick glass
plates under the cloak to minimize distortion in the image.

Cloaking in the experiment of Figs.~\ref{fig:setup} and
\ref{fig:green}(a) is proven as follows: if the cloak can hide the
wedge successfully, the CCD camera should capture an erect and
undeformed ``MIT'' as if there were nothing on top of the mirror. We
chose an incident angle $\theta = 18^{\circ}$ inside the tank with
respect to the horizontal reference plane (the external angle was
adjusted in the experiment to match the nominal value of $\theta$
according to Snell's law.) Figure~\ref{fig:green}(b) shows the image
when the wedge only without the cloak was placed on top of the
mirror. The letter ``M'', after being reflected on the wedge, is
located very far away from ``IT'' and misses the CCD.
Figure~\ref{fig:green}(c) shows the image when there is a flat
ground plane on top of the wedge directly. We can see that ``M''
itself is undistorted, but it is shifted upwards compared to ``IT'',
for the same reason as in Fig.~\ref{fig:rays2}(c).
Figure~\ref{fig:green}(d) shows the image when the cloak is placed
above the wedge. All letters in the captured image are located at
the same altitude, as if the cloaked wedge were not there. To make
sure the position where the light is reflected from the bottom of
the cloak did not influence the cloaking effect, we moved the
position of the cloak along the $z$ axis. No obvious extra
distortion was observed except when the reflection occurred across
the conjunction line between the two pieces of calcite crystal.

The flat ground plane for Fig.~\ref{fig:green}(c) was implemented as
follows: first we put a 2mm--tall flat steel plate close to the
wedge. We then placed a 1.5mm--thick piece of glass with one side
coated with chrome on top with the coated side touching both the
wedge's peak and the flat plate. Since the refractive index of glass
was almost perfectly matched to the laser liquid environment, this
thickness had negligible influence on the experimental results.

\begin{figure}
\begin{centering}
\includegraphics[width=1\columnwidth,draft=false]{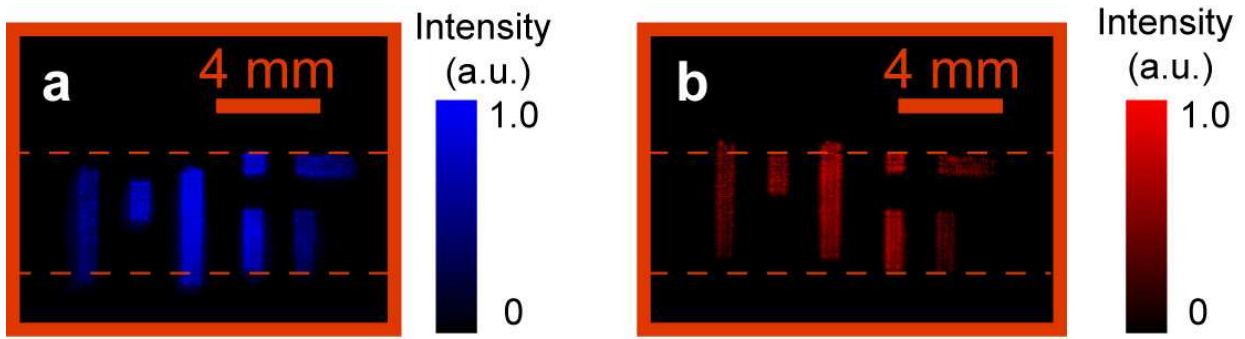}% Here is how to import EPS art
\caption{\label{fig:bluered} Images captured on the CCD camera
when the cloak is covering the wedge and the color of light is
changed to (a) blue at a wavelength of 488 nm, and (b) red at
a wavelength of 650 nm. The incident
angle $\theta = 18^{\circ}$ inside the tank.}
\end{centering}
\end{figure}

After the effectiveness of the cloak was tested under green light
(561 nm), the color was switched to blue (488 nm) and red (650 nm),
respectively. The resulting images in Fig.~\ref{fig:bluered} show
that the cloaking performance is still reasonably good. Because of
the color aberration, there was a shift of 450 $\mu$m at blue and
400 $\mu$m at red, respectively, in the image of ``M.'' The shift
was less than 10\% of the height of ``M'' (4.5mm). Further reduction
of the color aberration can be attained using methods inspired by
complementary dispersion~\cite{lensabberation}, which is out of the
scope of this Letter.

\begin{figure}
\begin{centering}
\includegraphics[width=0.8\columnwidth,draft=false]{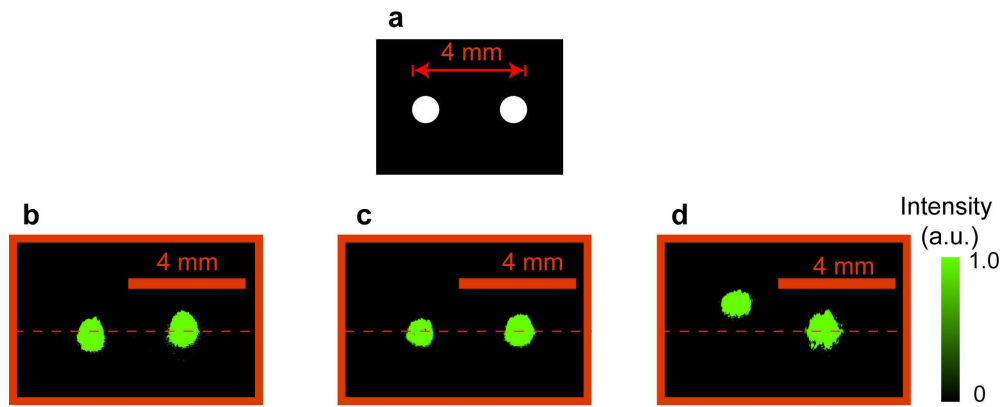}
\caption{\label{fig:twopoints} Images of (a) two spots with
incident angles inside the tank (b) $\theta=0^\circ$, (c) $\theta=18^\circ$ and (d) $\theta=40^\circ$.}
\end{centering}
\end{figure}

To verify quantitatively that our cloak functions independently of
incident angle, we used a simpler two-spot pattern
[Fig.~\ref{fig:twopoints}(a)] instead of the ``MIT'' pattern. The
illumination wavelength was again 561nm in this case. Performance
was verified for $\theta=0^\circ$ and $\theta=18^\circ$
[Figs.~\ref{fig:twopoints}(b) and~\ref{fig:twopoints}(c)] but
degraded somewhat for $\theta=40^\circ$
[Fig.~\ref{fig:twopoints}(d)] because at so large a reflected angle,
the error in the cloak dimensions was magnified. It is worth
mentioning that at $0^{\circ}$ incidence, the light did not actually
touch the bottom of the cloak or the mirror's surface. This
corresponds to a test of complete ``seeing through.''

The main limitation of our cloak currently is that it can only work
in a 2D geometry  (the light must propagate in the plane defined in
Fig.~1) and for only one polarization. However, it is possible to
extend this cloak to a three-dimensional (3D) geometry, just as the
2D carpet cloak~\cite{li_carpet,valentine,gabrielli,park} can be
extended to the 3D ``carpet'' cloak~\cite{ergin,huifeng}. Since it
is well known that the light transmitted underwater is mainly
polarized with the magnetic field parallel to the ground, our design
can be used in similar aqueous environments. The application of
transformation optics design in conventional optical fabrication
offers a cost effective yet accurate solution for making
invisibility cloaks. We believe that this technique will open up the
possibility of taking more transformation optics devices from
concepts into practical applications.

We acknowledge financial support from Singapore's National Research
Foundation and the National Institutes of Health.

\newpage

\pagebreak

\end{document}